\documentclass[aps,twocolumn,longbibliography,english,prx]{revtex4-1}
\usepackage[colorlinks=true,urlcolor=blue,citecolor=blue,linkcolor=blue]{hyperref} 
\usepackage[T1]{fontenc}
\usepackage{amssymb}
\usepackage{graphicx}
\usepackage{amsmath,color}
\usepackage{mathrsfs}
\usepackage{float}
\usepackage[normalem]{ulem}
\usepackage{indentfirst}
\usepackage{txfonts}

\emergencystretch=\maxdimen
\makeatletter

\def\journal #1, #2, #3, 1#4#5#6{{\sl #1~}{\bf #2}, #3 (1#4#5#6) }

\DeclareMathOperator*{\argmax}{arg\,max}

\newcommand{\<}{\langle}
\renewcommand{\>}{\rangle}

\renewcommand{\v}[1]{{\bf #1}}
\newcommand{\dataset}{{\mathcal{D}}}
\newcommand{\wfunc}{{\psi}}
\newcommand{\thetav}{{\boldsymbol{\theta}}}
\newcommand{\gammav}{{\boldsymbol{\gamma}}}
\newcommand{\thetai}{{\theta^\alpha_l}}
\newcommand{\Expect}{{\mathbb{E}}}

\newcommand{\xset}{\mathbf{X}}
\newcommand{\pdata}{\mathbf{\pi}}

\newcommand{\gammaset}{\boldsymbol{\Gamma}}
\newcommand{\ei}{{\mathbf{e}_l^\alpha}}
\newcommand{\sigmag}{{\nu}}
\newcommand{\BAS}{Bars-and-Stripes }

\newcommand{\qexpect}[1]{{\left\langle #1\right\rangle}}
\newcommand{\expect}[2]{{\mathop{\mathbb{E}}\limits_{\substack{#2}}\left[#1\right]}}
\newcommand{\pshift}[1]{{p_{\thetav+#1}}}

\newcommand{\Eq}[1]{Eq.~(\ref{#1})}
\newcommand{\Fig}[1]{Fig.~\ref{#1}}
\newcommand{\Ref}[1]{Ref.~\onlinecite{#1}}
\newcommand{\Sec}[1]{Sec.~\ref{#1}}
\newcommand{\App}[1]{Appendix \ref{#1}}

\newcommand{\material}[1]{\iffalse[{\bf  \color{cyan}{Material: #1}}]\fi}

\makeatother

\usepackage{babel}

\begin{document}

\title{Differentiable Learning of Quantum Circuit Born Machine}

\author{Jin-Guo Liu$^{1}$}
\author{Lei Wang$^{1,2}$}
\email{wanglei@iphy.ac.cn}

\affiliation{$^{1}$Institute of Physics, Chinese Academy of Sciences, Beijing 100190, China}
\affiliation{$^{2}$CAS Center for Excellence in Topological Quantum Computation, University of Chinese Academy of Sciences, Beijing 100190, China}

\begin{abstract}
    Quantum circuit Born machines are generative models which represent the probability distribution of classical dataset as quantum pure states. Computational complexity considerations of the quantum sampling problem suggest that the quantum circuits exhibit stronger expressibility compared to classical neural networks. One can efficiently draw samples from the quantum circuits via projective measurements on qubits. However, similar to the leading implicit generative models in deep learning, such as the generative adversarial networks, the quantum circuits cannot provide the likelihood of the generated samples, which poses a challenge to the training. We devise an efficient gradient-based learning algorithm for the quantum circuit Born machine by minimizing the kerneled maximum mean discrepancy loss. 
We simulated  generative modeling of the \BAS dataset and Gaussian mixture distributions using deep quantum circuits. Our experiments show the importance of circuit depth and gradient-based optimization algorithm. The proposed learning algorithm is runnable on near-term quantum device and can exhibit quantum advantages for generative modeling. 
\end{abstract}
\maketitle

\section{Introduction}
Unsupervised generative modeling is at the forefront of deep learning research~\cite{Goodfellow2016}. Unlike the extremely successful discriminative tasks such as supervised classification and regression, the goal of generative modeling is to model the probability distribution of observed data and generate new samples accordingly. Generative modeling finds wide applications in computer vision~\cite{Zhu2017}, speech synthesis~\cite{Van2016}, as well as chemical design~\cite{Gomez2016}. And it is believed to be a crucial component towards artificial general intelligence. However, generative modeling is more challenging than discriminative tasks since it requires one to efficiently represent, learn and sample from high-dimensional probability distributions~\cite{Goodfellow2016}.

In parallel to the rapid development of deep learning,
there is heated ongoing research to fabricate intermediate scale quantum circuits~\cite{Preskill2018,IBM2017,*Google2018}. Quantum hardware may show greater potential than their classical counterparts in generative tasks. Multiple schemes have been proposed to boost the performance of classical generative models using quantum devices. For example, quantum Boltzmann machines~\cite{Amin2016,Khoshaman2018,Benedetti2017} generalize the energy function of classical Boltzmann machines~\cite{Ackley1985} to quantum Hamiltonian for possible stronger representational power and faster training. 
A concern which may prevent the quantum Boltzmann machine from surpassing its classical counterpart is the limited connectivity on the actual quantum hardware~\cite{Dumoulin2014}.
\Ref{Gao2017} introduces a quantum generalization of the probabilistic graphical model~\cite{Koller2009},
which can be exponentially more powerful than its classical counterpart and has exponential speedup in training and inference at least for some instances under reasonable assumptions in computational complexity theory.

Another class of arguably simpler quantum generative models named Born machines~\cite{Han2017, Cheng2017, Benedetti2018} directly exploit the inherent probabilistic interpretation of quantum wavefunctions~\cite{Born1926}. Born machines represent probability distribution using quantum pure state instead of the thermal distribution like the Boltzmann machines~\cite{Ackley1985}. Therefore, Born machines can directly generate samples via projective measurement on the qubits, in contrast to the slow mixing Gibbs sampling approach. 
Moreover, computational complexity considerations on quantum sampling problems suggest that a quantum circuit can produce probability distribution that is \#P-hard~\cite{Fortnow1999,Aaronson2011,Lund2017},
which is infeasible to simulate efficiently using classical algorithms. The same reasoning underlines the current efforts towards "quantum supremacy" experiments by sampling outcomes of random quantum circuits~\cite{Boixo2016}. \Ref{Han2017} performed a classical simulation of the Born machine using the matrix product state representation of the quantum state. 
It will be even more promising to realize the Born machines using quantum circuits since one can at least efficiently prepare some of the tensor networks on a quantum computer~\cite{Huggins2018}. Recently, Ref.~\cite{Benedetti2018} demonstrated experimental realization of the Born machine on a four qubits shallow quantum circuit trained by gradient-free optimization of measurement histograms. 

To further scale up the quantum circuit Born machine (QCBM) to larger number of qubits and circuit depth, one needs to devise an appropriate objective function for the generative tasks without explicit reference to the model probability. Unlike the tensor network simulation~\cite{Han2017}, QCBM belongs to \emph{implicit} generative models since one does not have access to the wavefunction of an actual quantum circuit. Thus, QCBM can be used as a simulator to generate samples without access to their likelihoods, which is similar to the notable generative adversarial networks (GAN)~\cite{Goodfellow2014,Goodfellow2016c}. Compared to generative models with explicit likelihoods such as the Boltzmann machines~\cite{Hinton2006}, normalizing flows~\cite{Oord2016,Dinh2016,Kingma2016,Rezende2015,Papamakarios2017}, and variational autoencoders~\cite{Kingma2013,Rezende2014}, the implicit generative models can be more expressive due to less restrictions in their network structures. On the other hand, having no direct access to the output probability also poses challenge to the scalable training of quantum circuits. 


Moreover, one also needs better learning algorithm than the gradient-free optimization scheme~\cite{Benedetti2018}, especially given the noisy realization of current quantum circuits.
Similarly, scalability of the optimization scheme is also a crucial concern in deep learning, in which deep neural networks can even reach billions of parameters~\cite{Simonyan2014}. In the history of machine learning, gradient-free algorithms were employed to optimize small-scale neural networks~\cite{MinskyPerceptrons}. However, they failed to scale up to a larger number of parameters. It is the back-propagation algorithm~\cite{Rumelhart1986} which can efficiently compute the gradient of the neural network output with respect to the network parameters enables scalable training of deep neural nets. It is thus highly demanded to have scalable quantum algorithms for estimating gradients on actual quantum circuits. 


Recently, gradient-based learning of quantum circuits has been devised for quantum control~\cite{Li2017} and discriminative tasks~\cite{Farhi2018,Mitarai2018}. 
Although they are still less efficient compared to the back-propagation algorithm for neural networks, 
these unbiased gradient algorithms can already greatly accelerate the quantum circuit learning. Unfortunately, direct application of these gradient algorithms~\cite{Li2017,Farhi2018,Mitarai2018} to QCBM training is still non-trivial since the output of the generative model is genuinely bit strings which follow high-dimensional probability distributions. 
In fact, it is even an ongoing research topic in deep learning to perform differentiable learning of implicit generative model with discrete outputs~\cite{Goodfellow2016c, Li2017b}.  


In this paper, we develop an efficient gradient-based learning algorithm to train the QCBM. In what follows, we first present a practical quantum-classical hybrid algorithm to train the quantum circuit as a generative model in \Sec{sec-model}, thus realize a Born machine.
Then we apply the algorithm on $3\times 3$ \BAS and double Gaussian peaks datasets in \Sec{sec-app}.
We show that the training is robust to moderate sampling noise, and is scalable in circuit depth. Increasing the circuit depth significantly improves the representational power for generative tasks.
Finally, we conclude and discuss caveats and future research directions about the QCBM in \Sec{sec-discussion}.

\section{Model and Learning algorithm}\label{sec-model}
%

\begin{figure}
    \begin{center}
        \includegraphics[width=\columnwidth,trim={0.5cm 1.cm 0.5cm 1cm},clip]{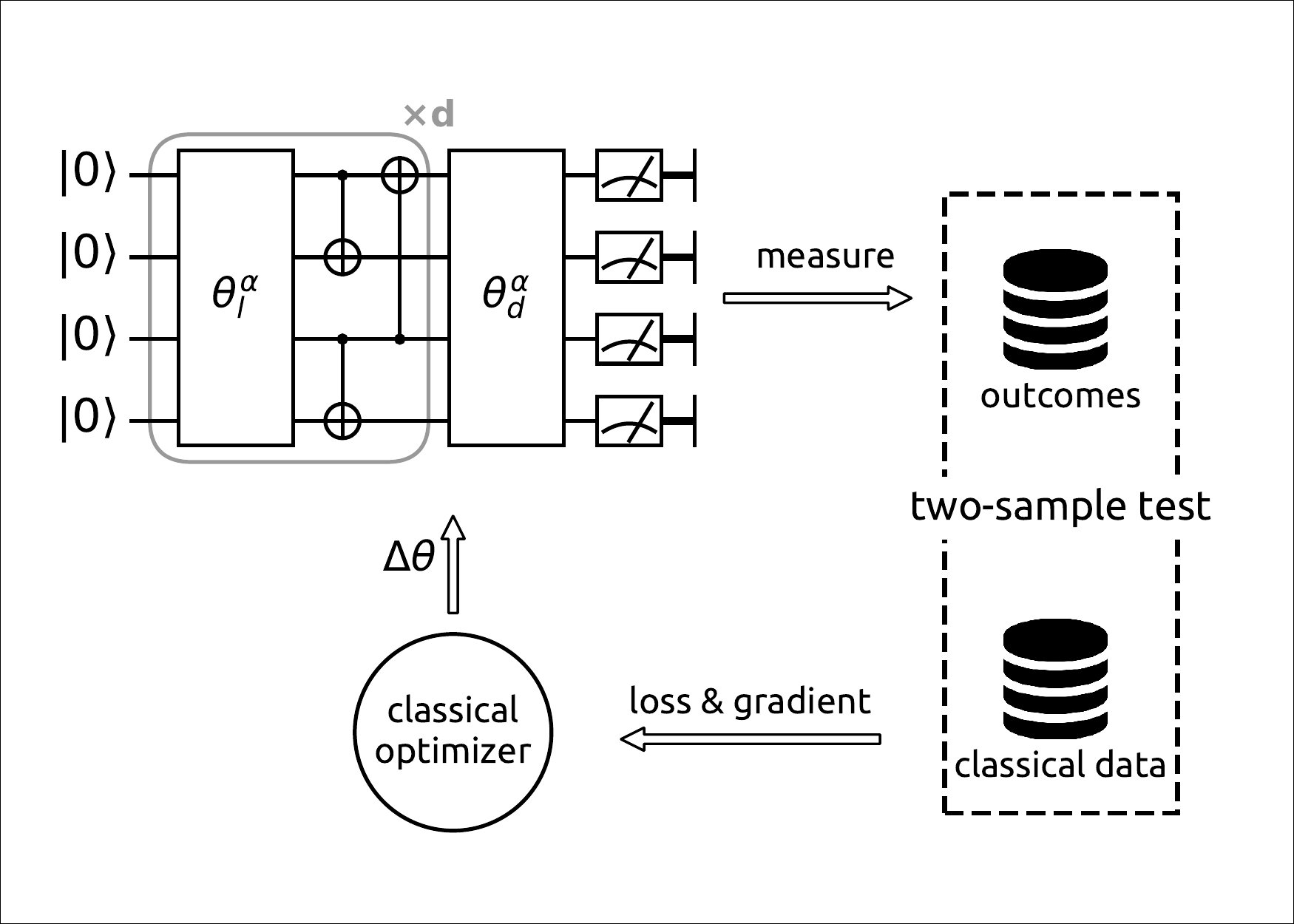}
    \end{center}
    \caption{Illustration of the differentiable QCBM training scheme.
    Top left is the quantum circuit which produce bit string samples.
    The dashed box on the right denotes two-sample test on the generated samples and training samples, with the loss function (\Eq{eq-mmd-loss}) and corresponding gradients (\Eq{eq-mmd-grad}) as outputs.
    $\Delta\thetav$ is the amount of updated to be applied to the circuit parameters, which are computed by a classical optimizer. The outcome of the training is to produce a quantum circuit which generates samples according to the learned probability distribution on the computational basis.} 
    \label{fig-concept}
\end{figure}

Given a dataset $\dataset=\{x\}$ containing independent and identically distributed (i.i.d.) samples from a target distribution $\pi(x)$, we set up a QCBM to generate samples close to the unknown target distribution. As shown in \Fig{fig-concept}, the QCBM takes the product state $|0\>$ as an input and evolves it to a final state $|\wfunc_\thetav\>$ by a sequence of unitary gates. Then we can measure this output state on computation basis to obtain a sample of bits $x\sim p_\thetav(x)=|\< x|\wfunc_\thetav\>|^2$. The goal of the training is to let the model probability distribution $p_\thetav$ approach to $\pdata$.  

We employ a classical-quantum hybrid feedback loop as the training strategy. The setup is similar to the Quantum Approximate Optimization Algorithm (QAOA)~\cite{Farhi2014,Farhi2017,Otterbach2017} and the Variational Quantum Eigensolver (VQE)~\cite{Peruzzo2014,OMalley2016, Kandala2017}.
By constructing the circuits and performing measurements repeatedly we collect a batch of samples from the QCBM. Then we introduce two-sample test as a measure of distance between generated samples and training set, which is used as our differentiable loss.
Using a classical optimizer which takes the gradient information of the loss function, we can push the generated sample distribution towards the target distribution. 

\subsection{Quantum Circuit Architecture Design}\label{sec:structurelearning}
The overall circuit layout is similar to the IBM variational quantum eigensolver \cite{Kandala2017}, where one interweaves single qubit rotation layers and entangler layers shown in \Fig{fig-concept}. 
The rotation layers are parameterized by rotation angles $\thetav=\{\thetai\}$,
where the layer index $l$ runs from $0$ to $d$, with $d$ the maximum depth of the circuit. 
$\alpha$ is a combination of qubit index $j$ and arbitrary rotation gate index, where the arbitrary rotation gate has the form $U(\theta_l^j)=R_z(\theta_l^{j,1})R_x(\theta_l^{j,2})R_z(\theta_l^{j,3})$ with $R_m(\theta)\equiv \exp\left(\frac{-i\theta\sigma_m}{2}\right)$.
The total number of parameters in this QCBM is $(3d+1)n$, with $n$ the number of qubits~ \footnote{The number of parameters here is not $3(d+1)n$ here because the leading and trailing $R_z$ gates can be omitted without affecting the  output probability distribution.}.
We employ CNOT gates with no learnable parameters for the entangle layers to induce correlations between qubits. In light of experimental constraints on the connectivity of the circuits, we make the connection of the entangle layers to be sparse by requiring its topology as a tree (i.e. the simplest connected graph). From the classical probabilistic graphical model's perspective~\cite{Koller2009}, the tree graph that captures information content of the dataset most efficiently is Chow-Liu tree~\cite{Chow1968}. Since controlled unitary gates have a close relation with classical probability graphical models~\cite{Low2014}, we employ the same Chow-Liu tree as the topology of CNOT gates. To construct the Chow-Liu tree we first compute mutual information between all pairs of the bits for samples in the training set as weights, and then construct the maximum spanning tree
using, for example, the Kruskal's algorithm.
The assignment of the control bit and the target bit on a bond is random, since the Chow-Liu algorithm treated directed and undirected graphs the same.
In the case where this connection structure is not directly supported by the hardware, a combination of SWAP gates and CNOT gates can be used to efficiently simulate the required structure~\cite{Zulehner2017}.
\material{
\begin{align}
    \begin{split}
        &T_{\rm Chow-Liu} = \argmax\limits_T\sum\limits_{(s,t)\in T}\mathbb{I}(s,t)\label{eq-log-likelihood},
    \end{split}
\end{align}
where the summation over $(s,t)\in T$ iterates over all edges in a tree graph, and $\mathbb{I}(s,t)$ is the two site mutual information between two bits in dataset, which is calculated using empirical distributions.
}

The performance of entangle layers constructed in this procedure is better than most random connections with the same number of gates. This data-driven quantum circuit architecture design scheme respects the information content of the classical dataset easier and may alleviate issues of vanishing gradients for large-scale applications~\cite{Huggins2018}.

\subsection{Loss Function and Gradient-based Optimization}
Viewing the QCBM as an implicit generative model~\cite{Mohamed2016},
we train it by employing the kernel two-sample test~\cite{Anderson1994}.
The idea is to compare the distance in the kernel feature space on the samples drawn from the target and the model distributions. We refer the following loss function as the squared maximum mean discrepancy (MMD)~\cite{Gretton2007,Gretton2012}
\begin{align}
    \begin{split}
        \mathcal{L} =&  \left\|\sum_{x} p_\thetav(x) \phi(x)- \sum_{x} \pi(x) \phi(x)  \right\|^2 \\
        =&\expect{K(x,y)}{{x\sim p_\thetav, y\sim p_\thetav } }-2\expect{K(x,y)}{x\sim p_\thetav,y\sim \pdata}+\expect{K(x, y)}{x\sim \pdata,y\sim \pdata}.
    \end{split}\label{eq-mmd-loss}
\end{align}
The summation in the first line runs over the whole Hilbert space. The expectation values in the second line are for the corresponding probability distributions. 
The function $\phi$ maps $x$ to a high-dimensional reproducing kernel Hilbert space~\cite{Hofmann2008}. However, as common in the kernel tricks, by defining a kernel function $K(x,y)=\phi(x)^T\phi(y)$ one can avoid working in the high-dimensional feature space. We employ a mixture of Gaussians kernel $K(x,y) = \frac{1}{c}\sum\limits_{i=1}^c\exp\left(-\frac{1}{2\sigma_i}|x-y|^2\right)$ to reveal differences between the two distributions under various scales. Here, $\sigma_i$ is the bandwidth parameter which controls the width of the Gaussian kernel.
The sample $x$ can either be a bit string (vector) or an integer (scalar) depending on the representation. When $x$ is a bit string, $|x|$ stands for the $\ell^2$-norm in the vector space. 
The MMD loss with Gaussian kernels asymptotically approaches zero if and only if the output distribution matches the target distribution exactly~\cite{Gretton2007,Gretton2012}.
The same loss function was used to train the generative moment matching networks (GMMN)~\cite{Li2015,Dziugaite2015,Li2017e}. 

To learn the QCBM as a generative model, we compute gradient of the loss function \Eq{eq-mmd-loss} with respect to the circuit parameters
\begin{eqnarray}
        \frac{\partial \mathcal{L}}{\partial \thetai} &=&\expect{K(x,y)}{x\sim p_{\thetav^+}, y\sim p_\thetav}-\expect{K(x,y)}{x\sim p_{\thetav^-},y\sim p_\thetav} \nonumber\\
        &-&\expect{K(x,y)}{x\sim p_{\thetav^+},y\sim \pdata}+\expect{K(x,y)}{x\sim p_{\thetav^-},y\sim \pdata}.\label{eq-mmd-grad}
\end{eqnarray}
Here, $p_{\thetav^+}(x)$ and $p_{\thetav^-}(x)$ are output probabilities of QCBM under circuit parameters $\thetav^{\pm}=\thetav\pm\frac{\pi}{2}\ei$, where $\ei$ is the $(l,\alpha)$-th unit vector in parameter space (i.e. $\thetai\leftarrow\thetai\pm\frac{\pi}{2}$, with other angles unchanged).
In contrast to the finite difference methods like simultaneous perturbation stochastic approximation (SPSA)~\cite{Spall1998}, \Eq{eq-mmd-grad} is an unbiased estimator of the exact gradient. Its detailed derivations is in \App{sec-mmd}.

In order to estimate the gradient [\Eq{eq-mmd-grad}] on an actual quantum circuit, one can repeatedly send rotation and entangle pulses to the device according to the circuit parameters $\thetav^{(\pm)}$, and then perform projective measurements on the computational basis to collect binary samples $x\sim p_{\thetav^{(\pm)}}$. While for $x\sim\pdata$, one can simply take a batch of data from the training dataset~\footnote{For the small dataset employed in our numerical experiment, we can afford to average over the whole training dataset.}. 
The sampling noise in the estimated gradient is controlled by the number of measurements $N$, denoted as the batch size. After one has obtained a sufficiently accurate gradient, one can use a classical optimizer to update the circuit parameters similar to the stochastic gradient descent training of deep neural nets. 

Parameter learning of quantum circuits is adaptive in the sense that the implementation of quantum gates can even be non-ideal. 
One can obtain gradients with high accuracy as long as the parametrized single qubits rotation gates are precise, which is relatively easier to achieve experimentally. The optimization scheme is independent of the detailed form of non-parametrized entangle gates.
Thus, the CNOT gate in the setup can be replaced by any gate which can generate desired quantum entanglements.

It is instructive to compare the training of the QCBM to that of classical implicit generative models such as GAN~\cite{Goodfellow2014,Goodfellow2016c} and GMMN~\cite{Li2015,Dziugaite2015,Li2017e}. Classically, one does not have access to the likelihood either. The gradient is thus obtained via the chain rule $\frac{\partial \mathcal{L}}{\partial \thetav}=\frac{\partial \mathcal{L}}{\partial x}\frac{\partial x}{\partial \thetav}$, which then does not apply for discrete data. 
On the other hand, the unbiased gradient estimator of the QCBM takes advantage of the  known structure of the unitary evolution and the MMD loss (see \App{sec-mmd}), despite that the probability of the outcome is unknown.  
In this sense, quantum circuits exhibit a clear quantum advantage over classical neural nets since they fill the gap of differentiable learning of implicit generative models of discrete data. 
\section{Numerical Experiments}\label{sec-app}

We carry out numerical experiments by simulating the learning of QCBM on a classical computer. These experiments reveal advantages of gradient based optimization over gradient-free optimization, and demonstrate stronger expressibility of deep circuits over shallow ones. 
The code can be found at the Github repository~\cite{github}

\subsection{\BAS Dataset}
\begin{figure}
    \begin{center}
        \includegraphics[width=\columnwidth,trim={0, 2cm 0 1cm}]{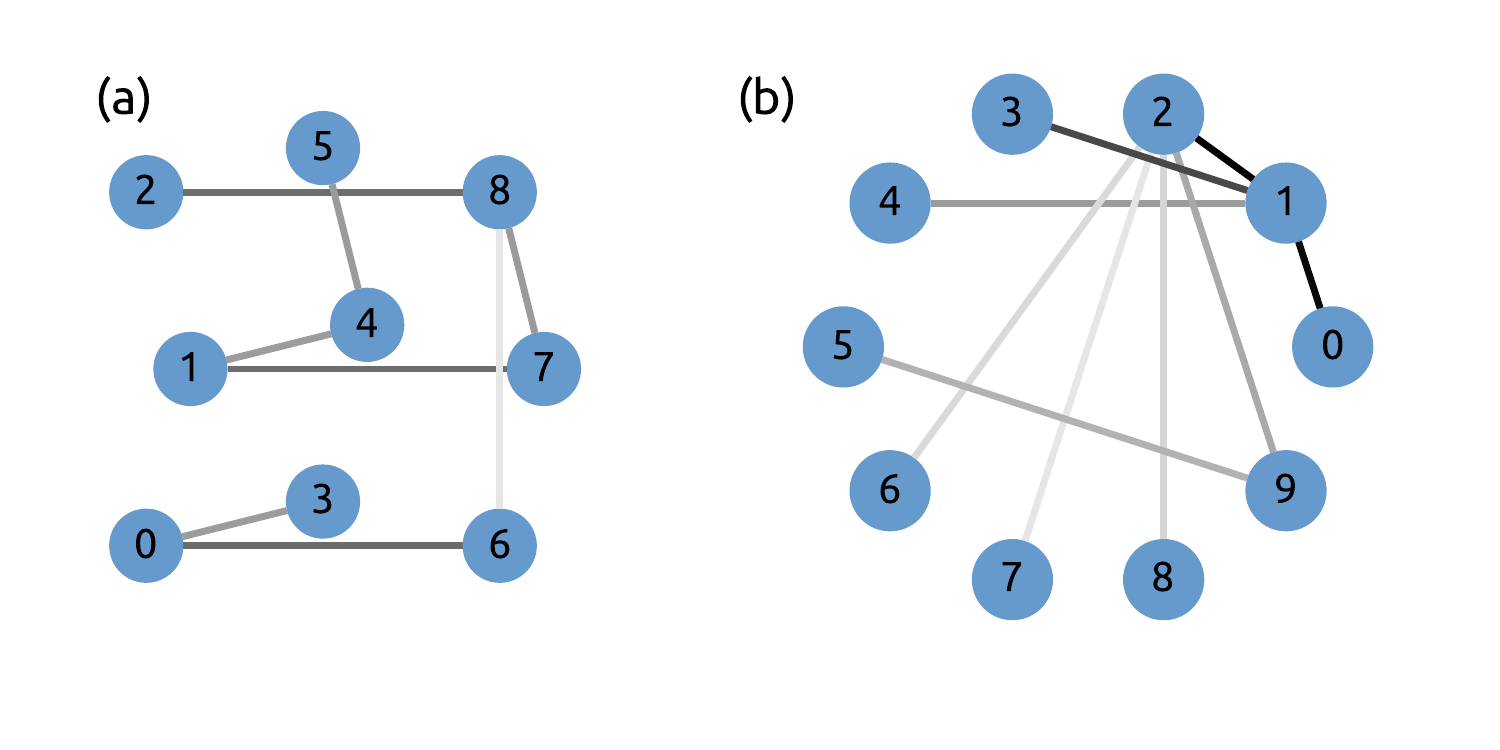}
    \end{center}
    \caption{(a) Connectivity of the CNOT gates for $3\times 3$ \BAS dataset generated via the chow-Liu tree algorithm. The qubits are arranged on a $3\times 3$ grid, with some of them shifted a bit in order to visualize the edges clearly.
    (b) Chow-Liu tree for double Gaussian peak model, the numbers represent the position of a bit in the digit, $0$ for the big end and $9$ for the little end.
    In this plot, the darkness of edges indicates the amount of mutual information between two sites.}\label{fig-chowliu-tree}
\end{figure}

We first train a QCBM on the \BAS dataset~\cite{Mackay2003,Han2017,Benedetti2018}, which is a prototypical image dataset consists of vertical bars and horizontal stripes. On a grid of $3\times 3$, the dataset contains 14 valid configurations. We model the pixels with a quantum circuit of $9$ qubits.
The Chow-Liu tree for this dataset is shown in \Fig{fig-chowliu-tree} (a).
Bonds are either row-wise or column-wise since correlations of pixels sharing the same row/column index are dominant in this dataset.
The bandwidths used in Gaussian kernels of MMD loss are $\sigma = 0.5,1,2,4$.

\begin{figure}
    \begin{center}
    \includegraphics[width=\columnwidth]{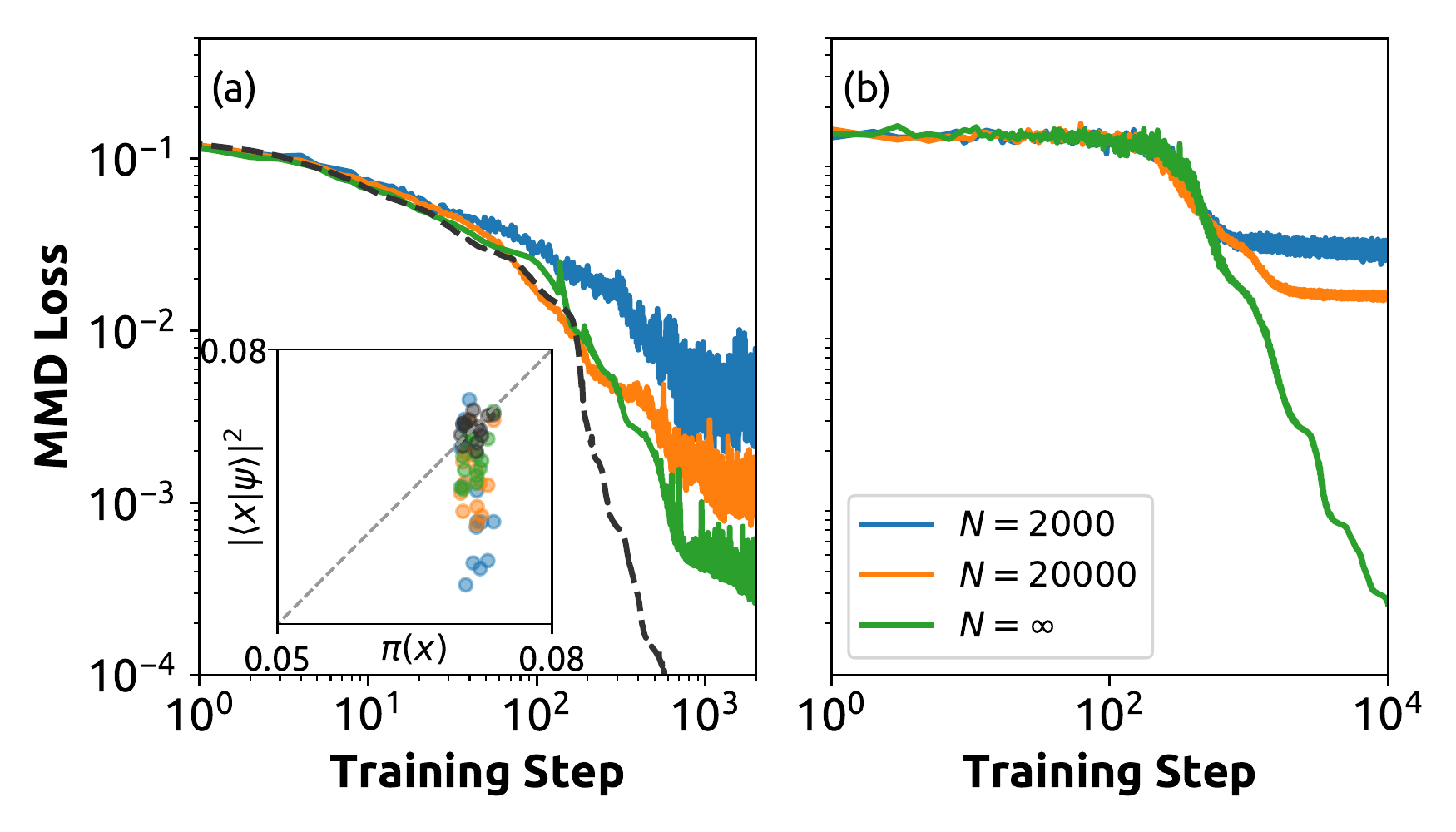}
    \end{center}
    \caption{
        The MMD loss (\Eq{eq-mmd-loss}) as a function of training steps under different sampling errors that governed by batch size $N$.
        (a) Gradient based training, solid colored lines are for Adam, and the dashed black line is for L-BFGS-B with $N=\infty$.
            Inset is a comparison of probability distribution between the training set and QCBM output, points on dashed line means exact match.
        (b) Gradient free CMA-ES training counterpart, where each point in the graph represents a mean loss of its population.
    }\label{fig-noiseloss}
\end{figure}

For circuit depth $d=10$, our gradient-based training is able to reduce the MMD loss efficiently. The loss function for different iteration steps is shown in \Fig{fig-noiseloss}(a).
We first perform L-BFGS-B~\cite{Byrd1995} optimization (black dashed line) using the exact gradient computed via the wavefunction ($N=\infty$) to test the expressibility of the quantum circuit. 
A loss of $2.4\times 10^{-7}$ can be achieved, showing that the circuit is quite expressive in terms of the two-sample test.  

In practice, one has to perform projective measurements on the qubits to collect statistics of the gradient since the wavefunction is inaccessible. This situation is similar to the mini-batch estimate of the gradient in deep learning~\cite{Goodfellow2016}. As is well known in the deep learning applications~\cite{Goodfellow2016}, the L-BFGS-B algorithm is not noise tolerant. Thus, it is unsuitable for quantum circuit learning in realistic situation. 
One needs to employ an alternate optimizer which is robust to the sampling noise to train the quantum circuit with noisy gradient estimator.

We employ the stochastic gradient optimizer Adam~\cite{Kingma2014} with the learning rate $0.1$. 
The sampling noise in the gradients can be controlled by tuning the batch size $N=2000,20000,\infty$ of the measurements. The solid lines in \Fig{fig-noiseloss} (a) show that as the sample size increases, the final MMD loss reduces systematically. 
The scatters in the inset confirmed that the model probability of learned quantum circuit and the target probability aligns better with lower MMD loss.

\begin{figure}
    \begin{center} 
    \includegraphics[width=\columnwidth,trim={0cm 0.5cm 1.5cm 1.5cm},clip]{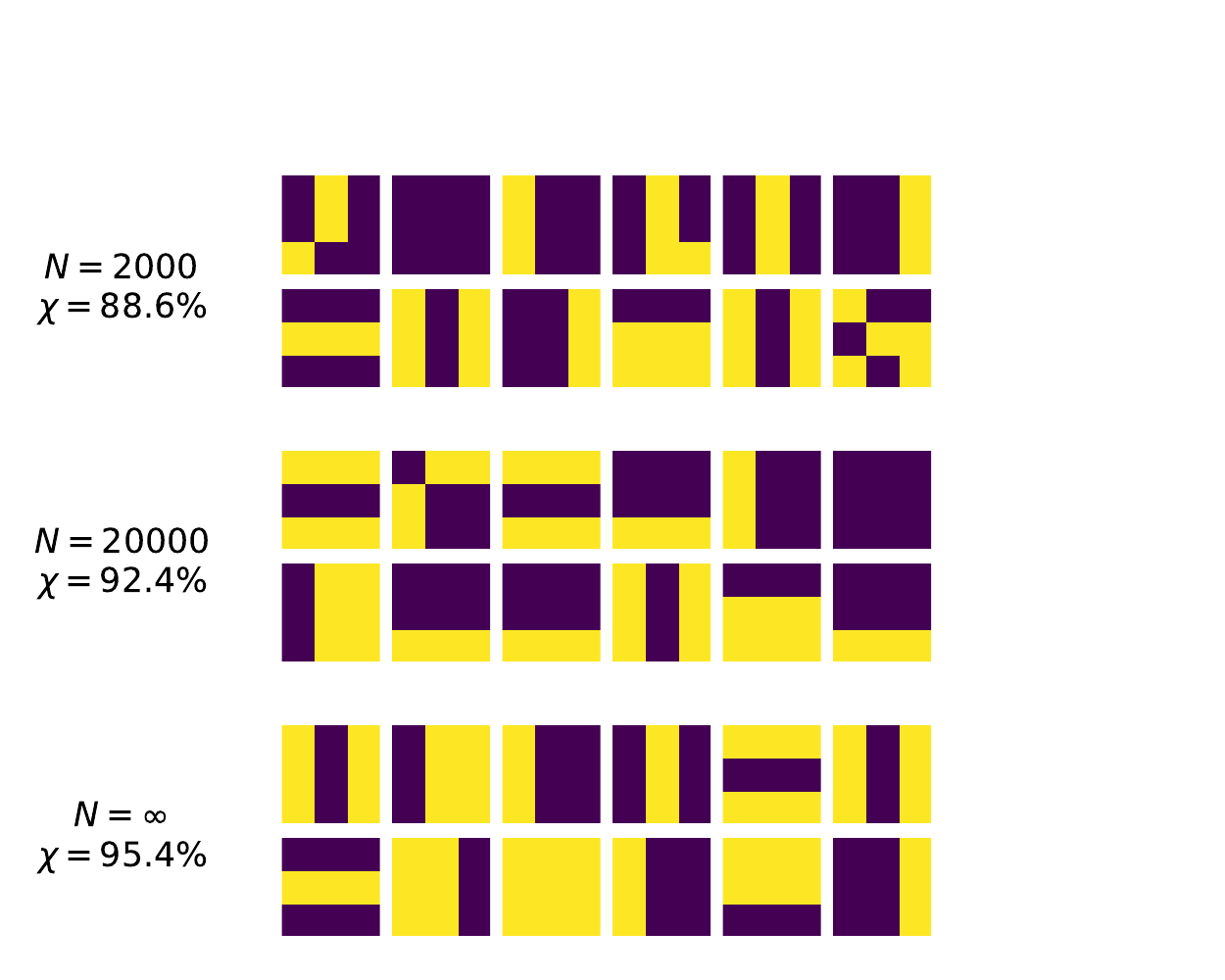}
    \end{center}
    \caption{$3\times3$ \BAS samples generated from QCBMs.
Circuit parameters used here are from the final stages of Adam training with different batch sizes $N$ in \Fig{fig-noiseloss} (a). $\chi$ is the rate of generating valid samples in the training dataset.
    For illustrative purpose, we only show $12$ samples for each situation with batch size $N$.}
    \label{fig-bs-gen} 
\end{figure}

To visualize the quality of the samples, we generated a few  samples from the QCBM trained under different measurement batch size $N$ in \Fig{fig-bs-gen}.
Here, we define a valid rate $\chi\equiv p(x \text{ is a bar or a stripe})$ as a measure of generation quality.
The valid rate increases as the batch size increases. 
However, even with a moderate number of measurement $N=2000$ one can achieve a valid rate $\chi=88.6\%$.
Here, we should mention that the best valid rate of $d=10$ layer circuit is achieved by L-BFGS-B optimizer with $N=\infty$, which is $\chi=99.9\%$.

To highlight the importance of using a gradient-based optimizer, we compare our approach to the covariance matrix adaptation evolution strategy (CMA-ES)~\cite{Hansen2006,Rios2013}, a state-of-the-art gradient-free stochastic optimizer.
The input of CMA-ES is the scalar loss function measured on the circuit instead of the vector gradient information. The CMA-ES optimizer is able to optimize non-smooth non-convex loss functions efficiently~\cite{Rios2013}, thus in general  performs better than other gradient-free methods such as the SPSA~\cite{Spall1998} in training noisy quantum circuits. We have confirmed this in our simulation.

In the absence of sampling noise $N=\infty$ in \Fig{fig-noiseloss} (b), we do observe that the CMA-ES optimizer is able to achieve similar performance as the Adam optimizer after $10^4$ steps of optimization with a population size of $50$. The total number of generated samples is $10^4\times50\times N$, which is comparable to the Adam training in \Fig{fig-noiseloss} (a)
\footnote{Notice for this circuit of depth $d=10$ that having $279$ parameters, we need to make $279\times N\times2+1$ measurements in a gradient estimation.}. 

However, the performance of CMA-ES deteriorates significantly once taking sampling noise into consideration, as shown for $N=2000$ and $N=20000$ in \Fig{fig-noiseloss} (b).
A possible explanation is that in each step of CMA-ES, its evolution strategy chooses the direction to go by inspecting the center of top $20\%$ instances.
This process can be understood as an effective finite difference gradient estimation base on the losses of its population. However, extracting gradient information from noisy losses is difficult, even one has plenty of them.


Another advantage of using gradient-based learning is the efficiency comparing with gradient-free methods, which gets particularly significant when circuits get deeper and the number of parameters increases.
In the following, we address the necessity of using deep circuits.
\begin{figure}
    \begin{center}
    \includegraphics[width=\columnwidth]{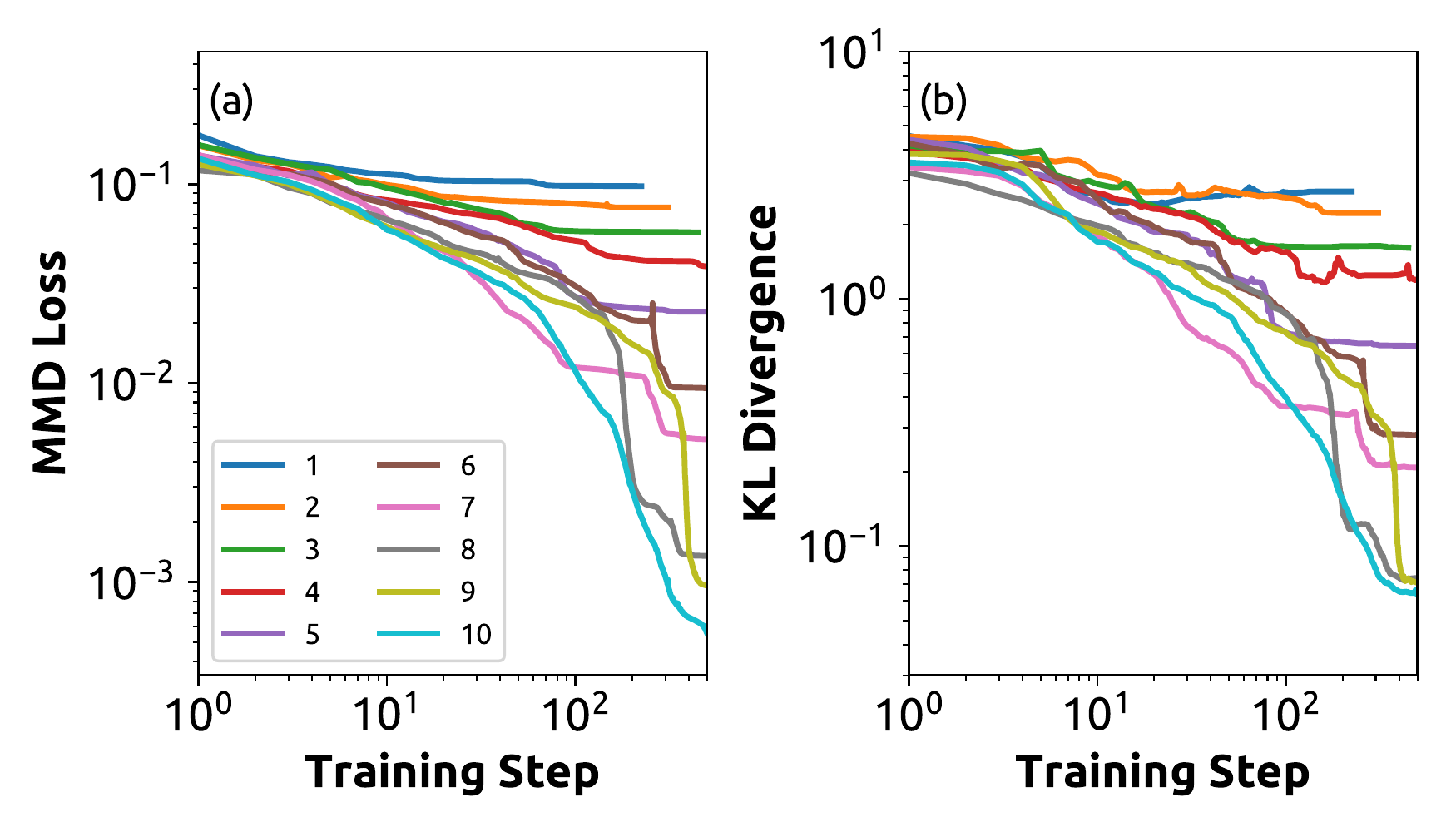}
    \end{center}
    \caption{Losses as a function of training step for circuit depth $d=1,\ldots,10$. (a) The MMD loss \Eq{eq-mmd-loss}, and (b) the corresponding KL divergence. Here, we use L-BFGS-B optimizer with exact gradient. 
     }\label{fig-mmd-kl}
\end{figure}
\Fig{fig-mmd-kl} (a) shows the MMD loss as a function of L-BFGS-B training steps for different circuit depth. One obtains lower loss for deeper quantum circuit after $500$ optimization steps. \Fig{fig-mmd-kl} (b) shows the Kullback-Leibler (KL) divergence~\cite{Kullback1951} calculated using the circuit parameters in (a) at different training steps. Note that this quantity is inaccessible for large-scale problems since one has no access to the target nor the output probability. We compute the KL divergence for the toy model to demonstrate that the MMD loss is a good surrogate for practical training.
The result indeed shows a consistency between MMD loss and KL divergence. And it also supports the observation that deep circuits have stronger representational power. 
Similar to deep neural networks, deep circuits can achieve better performance also due to that one is less prone to be trapped in a poor local minima with larger amount of parameters~\cite{Choromanska2015}. 

Another advantage of the QCBM over traditional deep neural networks is that its training not suffer from gradient vanishing/exploding problem as the circuit goes deeper. Gradient vanishing/exploding is a common problem for a traditional deep neural network~\cite{He2016} which originates from multiplications of a long chain of matrices in the back-propagation algorithm. Training of the deep quantum circuits naturally circumvented this problem by due to the unitary property of the time evolution. Similar idea was exploited in constructing classical recurrent neural networks with unitary building blocks~\cite{Jing2016}. More numerical simulation and analytical explanations can be found in \App{sec-gradient}.

\subsection{Mixture of Gaussians}
\begin{figure}
    \begin{center}
    \includegraphics[width=\columnwidth]{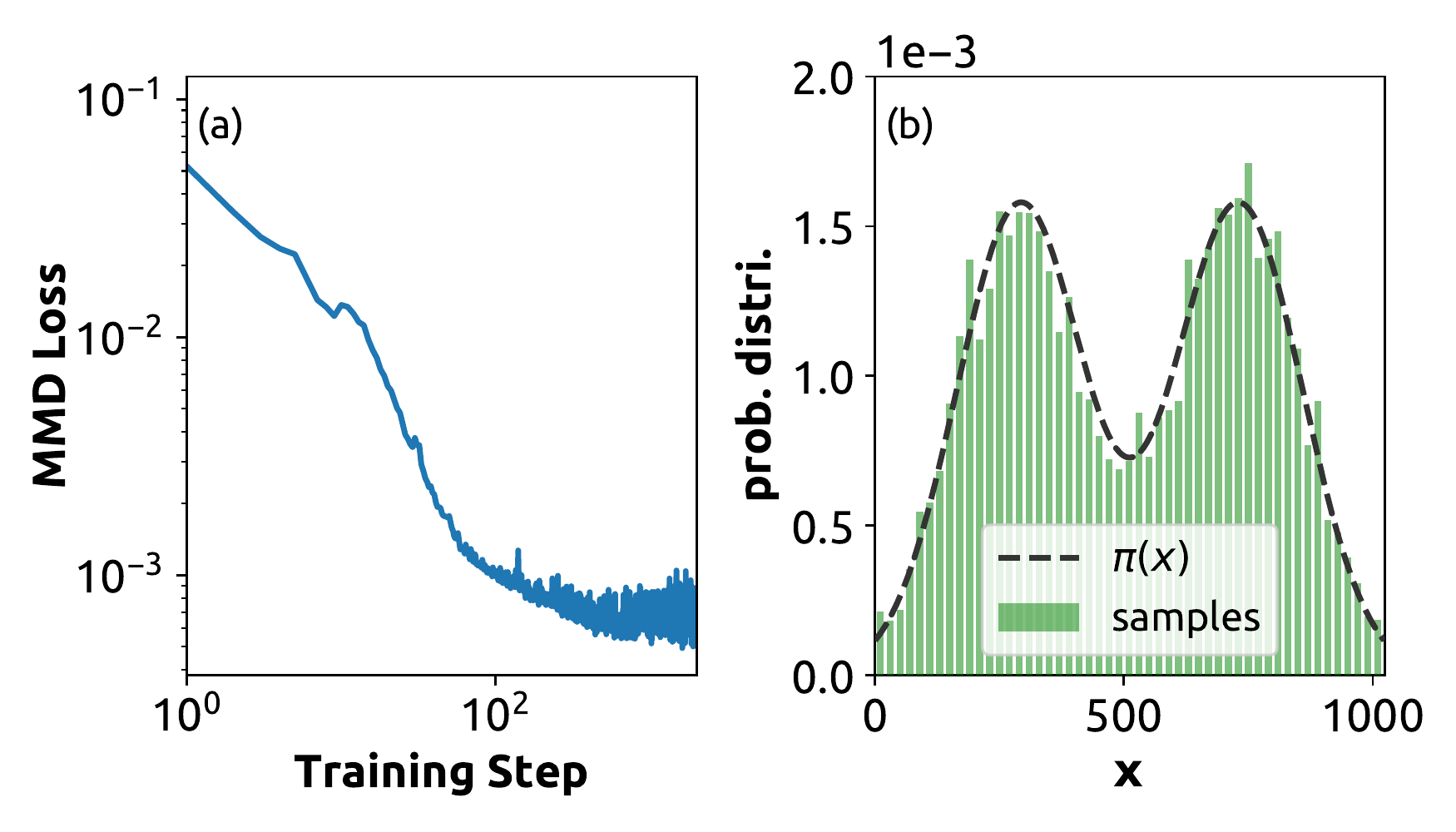}
    \end{center}
    \caption{(a) The MMD loss as a function of Adam training step.
    (b) Histogram for samples generated by a trained QCBM with a bin width $20$ (green bars),
    in comparison with the exact probability density function (black dashed line).}\label{fig-gaussian2}
\end{figure}

Next, we train a QCBM to model a mixture of Gaussians distribution 
\begin{equation}
    \pdata(x)\propto e^{-\frac{1}{2}\left(\frac{x-\mu_1}{\sigmag}\right)^2}+e^{-\frac{1}{2}\left(\frac{x-\mu_2}{\sigmag}\right)^2}. 
\end{equation}
Here, $x=1,\ldots,x_{\rm max}$ is an integer encoded by the qubits, with $x_{\rm max}=2^n$ and $n$ is the number of qubits. It is different from \BAS dataset, in which case a sample $x$ is represented as a bit string. We choose $\sigmag=\frac{1}{8}x_{\rm max}$,
the centers $\mu_1=\frac{2}{7}x_{\rm max}$ and $\mu_2=\frac{5}{7}x_{\rm max}$. The distribution is shown as the dashed line in \Fig{fig-gaussian2}(b).

In the following discussion, we use $n=10$ qubits and set circuit depth $d=10$.
Unlike the case of \BAS, the Gaussian mixture distribution is smooth and non-zero for all basis state.
Here, we generate $10^5$ i.i.d. samples from the target distribution as the training set.
Its Chow-Liu tree is shown in \Fig{fig-chowliu-tree}(b).
In this graph, we see the main contributions of mutual information are from bits near the big end (most significant bits labeled by small indices).
This is because the bit near the little end only determines the local translation of the probability on data axis. 
But for a smooth probability distribution, the value of the little end is nearly independent from values of the rest bits.
For example, the value of the big-end $0$-th bit being $0$/$1$ corresponds to the global left/right peak in \Fig{fig-gaussian2} (b). While the probability for the little end being $0/1$ corresponds to $x$ being even/odd. 

We use mixture of three Gaussian kernels with bandwidths $\sigma = 0.25, 10, 1000$.
$\sigma=0.25$ captures the local difference in distribution,
and $\sigma=1000$, which has the same scale as $x_{\rm max}$, which captures the overall differences in the probability distribution.

\Fig{fig-gaussian2} (a) shows the MMD loss as a function of the Adam optimization steps, with a sample size $N=20000$.
After $2000$ training steps, the MMD loss decreased from $9.6\times10^{-2}$ to $6.4\times10^{-4}$. To see whether this low MMD loss represents a good generative model, we then generate $20000$ samples from the QCBM and plot its binned histogram in \Fig{fig-gaussian2} (b).
We see an excellent match between the histogram (green bars) and the exact probability distribution (black dashed curve). Thus, we conclude that MMD loss with Adam optimizer can also learn a smooth probability distributions over the qubit index of a QCBM. Here, we acknowledge that the unbinned histogram appears more spiky, partly due to the MMD loss does not capture the local variation of the probability distribution.  Better circuit architecture design for representing continuous distribution may help alleviate this problem.

\section{Discussions}\label{sec-discussion}
We presented a practical gradient-based learning scheme to train quantum circuit Born machine as a generative model. The key component of the learning algorithm is to measure the gradient of the 
MMD two-sample test loss function \Eq{eq-mmd-grad} on a quantum computer unbiasedly and efficiently.
Besides possessing stronger representation power, the QCBM does not suffer from the gradient vanishing/exploding problem as circuit depth increases compared to the classical deep neural networks. While compared to other quantum generative models~\cite{Amin2016,Gao2017}, the quantum circuit Born machine has fewer restrictions on hardware and circuit design, while both the training and sampling can be efficiently 
carried out.


A recent work~\cite{Mcclean2018} pointed out that the quantum circuit also faces gradient vanishing problem as the number of qubits increases.
They found that the variance of gradient amplitudes decreases exponentially as the number of qubits increases in a random quantum circuit.
However, it is reasonable to believe that with better circuit structure design and parametrization strategy, such \Sec{sec:structurelearning} and Ref.~\cite{Huggins2018}, or using shared weights such as done in the convolutional neural networks~\cite{Goodfellow2016}, the gradient vanishing problem can be alleviated.
How gradient vanishing really affects gradient-based training in a large-scale QCBM needs further systematic investigation.

Our simulation of the quantum circuits Born machine is limited to a small number of qubits, thus the training set contains all patterns and we are pushing the circuits towards the memorization limit. In future applications, one shall focus on the generalization ability of the quantum circuits. In those cases, the structure and depth of a quantum circuit provide means of regularization since they can be design to express inductive bias in the natural distributions. In terms of the learning, the randomness in the stochastic gradient is also in favor of the generalization~\cite{Choromanska2015}.

Besides the two-sample test loss employed in this paper, one can explore alternative training schemes, e.g. adversarial
training~\cite{Goodfellow2014}. Alternatively, learning of the kernel function used in \Eq{eq-mmd-loss} may also improve the generation quality~\cite{Li2017e}. 
Finally, differentiable learning of the QCBM may be used for solving combinatorial optimization problems 
and structure learning tasks, where the outputs are encoded in discrete bit strings. 

\section{Acknowledgment}
Simulations of quantum circuits were mainly performed using the ProjectQ library~\cite{Steiger2016,Haner2018}.
We thank Alejandro Perdomo-Ortiz, Miles Stoudenmire, Damian Steiger, Xun Gao and Pan Zhang for helpful discussions. The authors are supported by the National Natural Science Foundation of China under the Grant No. 11774398 and research program of the Chinese Academy of Sciences under Grant No. XDPB0803. 

\appendix
\section{Unbiased gradient estimator of the probability of a quantum circuit}\label{sec-mmd}
\subsection{MMD Loss}
In the following, we derive \Eq{eq-mmd-grad} in the main text, starting from the partial derivative of \Eq{eq-mmd-loss}
\begin{align}
    \begin{split}
        \frac{\partial \mathcal{L}}{\partial \thetai} = &\sum\limits_{x,y} K(x,y) \left(p_\thetav(y)\frac{\partial p_\thetav(x)}{\partial \thetai}+p_\thetav(x)\frac{\partial p_\thetav(y)}{\partial \thetai}\right)\\
        &-2\sum\limits_{x,y} K(x,y) \frac{\partial p_\thetav(x)}{\partial \thetai}\pdata(y).
    \end{split}\label{eq-mmd-grad-raw}
\end{align}
The partial derivative on the right-hand side of this equation can be unbiasedly computed using the approach of \cite{Li2017, Mitarai2018}.
For a circuit containing a unitary gate parametrized as $U(\eta)=e^{-\frac{i}{2}\eta\Sigma}$ with $\Sigma^2=1$ (e.g. $\Sigma$ can be Pauli operators, CNOT or SWAP), the gradient of the expected value of an observable $B$ with respect to the parameter $\eta$ reads
\begin{equation}
    \frac{\partial \qexpect{B}_{\eta}}{\partial \eta}=\frac{1}{2}\left(\qexpect{B}_{\eta^+}-\qexpect{B}_{\eta^-}\right),\label{eq-opgrad}
\end{equation}
where $\qexpect{\cdot}_{\eta^{(\pm)}}$ represents expectation values of observables with respect to the output quantum wave function generated by the same circuit with circuit parameter $\eta^{\pm}\equiv\eta\pm\frac{\pi}{2}$. Note \Eq{eq-opgrad} is an unbiased estimate of the gradient in contrast to the finite difference approachs which are sensitive to noise of the quantum circuit.

Since the quantum circuit Born machine employs the same parametrization, we 
identify $|x\>\< x|$ as the observable and apply \Eq{eq-opgrad} to the output probability of the circuit
\begin{equation}
    \frac{\partial p_\thetav(x)}{\partial \thetai}=\frac{1}{2}\left(p_{\thetav^+}(x)-p_{\thetav^-}(x)\right),\label{eq-probdiff}
\end{equation}
where $\thetav^\pm\equiv\thetav\pm\frac{\pi}{2}\ei$, with $\ei$ the $(l,\alpha)$-th unit vector in parameter space. 
Substituting \Eq{eq-probdiff} into \Eq{eq-mmd-grad-raw} we have
\begin{align}
    \begin{split}
        \frac{\partial \mathcal{L}}{\partial \thetav}& =\frac{1}{2}\left(\sum\limits_{x,y} K(x,y) p_\thetav(y) p_{\thetav^+}(x)-\sum\limits_{x,y} K(x,y) p_\thetav(y) p_{\thetav^-}(x)\right)\\
        &+\frac{1}{2}\left(\sum\limits_{x,y} K(x,y) p_\thetav(x) p_{\thetav^+}(y)-\sum\limits_{x,y} K(x,y) p_\theta(x) p_{\thetav^-}(y)\right)\\
        &-\left(\sum\limits_{x,y} K(x,y) p_{\thetav^+}(x)\pdata(y)-\sum\limits_{x,y} K(x,y) p_{\thetav^-}(x)\pdata(y)\right).
    \end{split}
\end{align}
Using the symmetric condition of the kernel $K(x,y)=K(y,x)$, we arrive at \Eq{eq-mmd-grad} in the main text. This gradient algorithm for QCBM scales as $O(d^2)$ with the depth of the circuit, which is less efficient compared to the linear scaling back-propagation algorithm for classical neural networks. It is still unknown whether one can reach similar scaling on a quantum computer since the back-propagation algorithm requires cached intermediate results in the forward evaluation pass.
We have checked the correctness of the gradient estimator \Eq{eq-mmd-grad} against numerical finite difference. 

\subsection{Generalization to V-statistic}
To gain a better understanding of what kind of losses for a quantum circuit can be easily differentiated, we generalize the above result by considering an arbitrary function $f(\xset)$, with a sequence of bit strings $\xset\equiv\{x_1,x_2,\ldots, x_r\}$ as its arguments.
Let's define the following expectation of this function
\begin{equation}
    \Expect_f(\gammaset)\equiv\expect{f(\xset)}{\{x_i\sim \pshift{\gammav_i}\}_{i=1}^{r}}. 
\end{equation}
Here, $\gammaset=\{\gammav_1, \gammav_2,\ldots,\gammav_r\}$ is the offset angles applied to circuit parameters,
which means the probability distributions of generated samples is
$\{\pshift{\gammav_1}, \pshift{\gammav_2},\ldots ,\pshift{\gammav_r}\}$.
Writing out the above expectation explicitly, we have
\begin{equation}
    \Expect_f(\gammaset)=\sum\limits_\xset f(\xset)\prod\limits_i \pshift{\gammav_i}(x_i),
\end{equation}
where index $i$ runs from $1$ to $r$. Its partial derivative with respect to $\thetai$ is
\begin{equation}
    \frac{\partial \Expect_f(\gammaset)}{\partial \thetai}=\sum\limits_\xset f(\xset)\sum\limits_j\frac{\partial \pshift{\gammav_j}(x_j)}{\partial\thetai}\prod\limits_{i\neq j} \pshift{\gammav_i}(x_i)
\end{equation}
Again, using \Eq{eq-probdiff}, we have
\begin{align}
    \begin{split}
        \frac{\partial \Expect_f(\gammaset)}{\partial \thetai}&=\frac{1}{2}\sum\limits_{j,s=\pm}\sum\limits_\xset f(\xset){\pshift{\gammav_j+s\frac{\pi}{2}\ei}(x_j)}\prod\limits_{i\neq j} \pshift{\gammav_i}(x_i)\\
        &=\frac{1}{2}\sum\limits_{j,s=\pm}\Expect_f(\{\gammav_i+s\delta_{ij}\frac{\pi}{2}\ei\}_{i=1}^{r})
    \end{split}\label{eq-grad}
\end{align}
If $f$ is symmetric, $\Expect_f(\mathbf{0})$ becomes a V-statistic~\cite{Mises1947}, then \Eq{eq-grad} can be further simplified to
\begin{align}
    \frac{\partial \Expect_f(\gammaset)}{\partial \thetai}=\frac{r}{2}\sum\limits_{s=\pm}\Expect_f\left(\{\gammav_0+s\frac{\pi}{2}\ei,\gammav_1,\ldots,\gammav_r\}\right),
    \label{eq-grad-simple}
\end{align}
which contains only two terms. This result can be readily verified by calculating the gradient of MMD loss,
noticing the expectation of a kernel function is a V-statistic of degree $2$.
By repeatedly applying \Eq{eq-grad}, we will be able to obtain higher order gradients.

\subsection{Gradient of the KL-Divergence}
The KL divergence~\cite{Kullback1951} reads
\begin{equation}
    \mathbb{KL}(\pdata||p_\thetav)=-\sum\limits_{x}\pdata(x)\ln \left[{\frac {p_\thetav(x)}{\pdata(x)}}\right].
\end{equation}
Its gradient is
\begin{equation}
    \frac{\partial \mathbb{KL}}{\partial \thetai} = -\sum\limits_{x}\pdata(x)\frac{p_{\thetav^+}(x)-p_{\thetav^-}(x)}{2p_\thetav(x)},
\end{equation}
where the $\frac{p_{\thetav^+}(x)-p_{\thetav^-}(x)}{p_\thetav(x)}$ term cannot be transformed to a sampling problem on the quantum circuit. Since minimizing the negative log-likelihood is equivalent to minimizing the KL-divergence, it faces the same problem.

\section{Gradient Analysis}\label{sec-gradient}
\begin{figure}[!t]
    \begin{center}
    \includegraphics[width=\columnwidth,trim={0cm 0cm 0 0cm},clip]{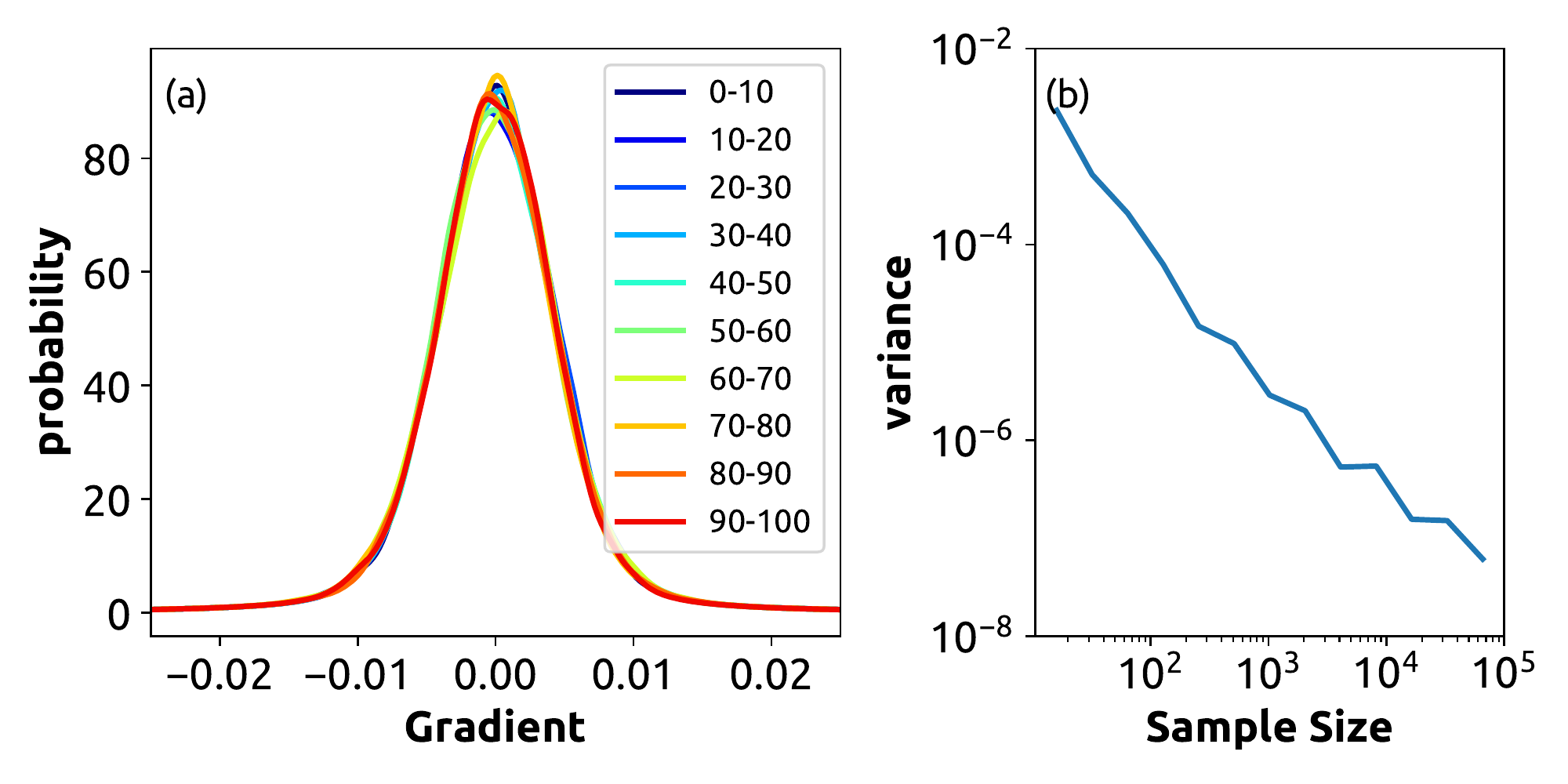}
    \end{center}
    \caption{(a) The probability distribution of gradient elements in different layers grouped by depth. The circuit parameters are chosen randomly.
        (b) The variance of MMD loss gradients as a function of sample size $N$.
    }\label{fig-grad-layer}
\end{figure}

In this appendix, we analyze the effect of depth and sampling error to gradients.
For a $9$ qubit quantum circuit of depth $d=100$ with random parameters
used in generating the \BAS dataset in the main text.
Histogram of MMD loss gradient values in different layers are shown in \Fig{fig-grad-layer} (a).
Distributions of gradients in different layers have no significant differences,
thus deep QCBM does not suffer from vanishing or exploding gradients when the layers go deeper.
This property has its physical origin in how the wave function changes when a perturbation is applied to $\thetai$ of the quantum circuit. 
Suppose we have a quantum circuit that performing the following evolution,
\begin{equation}
    |\wfunc\>=U_{N:1}|0\>
\end{equation}
Introduce a perturbation $\delta$ to $\thetai$,
\begin{equation}
    |\wfunc'\>=U_{N:k}e^{i\frac{\thetai+\delta}{2}\Sigma_k}U_{k-1:1}|0\>
\end{equation}
Its fidelity with respect to unperturbed wave function is
\begin{align}
    \begin{split}
        F&=|\<\wfunc|\wfunc'\>|\\
        &=|1-\frac{\delta^2}{8}+i\frac{\delta}{2}\<\phi|\Sigma_k|\phi\>|+O(\delta^3)\\
        &\simeq\sqrt{1-\frac{\delta^2}{4}(1-\<\phi|\Sigma_k|\phi\>^2)}
    \end{split}
\end{align}
Here, $|\phi\>\equiv U_{k-1:1}|0\>$ can be assumed to be random in a deep circuit. Thus we have the fidelity susceptibility~\cite{You2007,Wang2015} $\chi_F=\frac{\partial^2\ln F}{\partial \delta^2}=0.25(1-\qexpect{\Sigma_k}^2)$ which is independent of $\theta^k$.
Since we have $\Sigma_k^2=1$, $\chi_F$ is bounded by $0.25$, and is independent of the layer index. 

Next, we show the error in gradient estimation will decrease systematically as the number of samples increases for MMD loss.
As can be seen from the curve of variance calculated using $100$ copies of independent samples in \Fig{fig-grad-layer} (b).
In our case, to give a valid estimation of gradients, the standard error should be lower than the typical amplitude of gradients.
As can be referred from \Fig{fig-grad-layer} (a), the typical amplitude of gradient is approximately $5\times 10^{-3}$,
thus the sample size should be larger than $10^3$ in order to give a good estimation to these gradients.

\bibliographystyle{apsrev4-1}
\bibliography{LearnQcircuit.bib}

\end{document}